%% file: uqso_paper-arxiv.tex
\documentclass[12pt,preprint]{aastex}

\usepackage{amssymb,amsmath}
\usepackage{lineno}
\usepackage[]{natbib}
\usepackage{graphicx}

\usepackage{booktabs}
\usepackage{hyperref}
\hypersetup{%
    colorlinks=true,%
    citecolor=blue,%
    filecolor=blue,%
    linkcolor=blue,%
    urlcolor=blue
}
\usepackage{aas_macros}

\newcommand{\cygxi}{\mbox{Cyg\,X-1}}
\newcommand{\cygxiii}{\mbox{Cyg\,X-3}}

\newcommand{\cirxi}{\mbox{Cir\,X-1}}
\newcommand{\gx}{\mbox{GX\,339$-$4}}

\newcommand{\sss}{\mbox{SS\,433}}
\newcommand{\hhh}{\mbox{H1743$-$322}}
\newcommand{\xtejmdl}{\mbox{XTE\,J1550$-$564}}
\newcommand{\igrj}{\mbox{IGR\,J17091$-$3624}}
\newcommand{\eq}{Eq.}
\newcommand{\dd}{\; \mathrm{d}}

\begin{document}

\title{A Search for Time Dependent Neutrino Emission from Microquasars with the ANTARES Telescope}

\input{antares_authors_aastex.tex}

\begin{abstract}
Results are presented on a search for neutrino emission from a sample of six microquasars, based on the data collected by the ANTARES neutrino telescope between 2007 and 2010. By means of appropriate time cuts, the neutrino search has been restricted to the periods when the acceleration of relativistic jets was taking place at the microquasars under study. The time cuts have been chosen using the information from the X-ray telescopes RXTE/ASM and Swift/BAT, and, in one case, the gamma-ray telescope Fermi/LAT. Since none of the searches has produced a statistically significant signal, upper limits on the neutrino fluences are derived and compared to the predictions from theoretical models.
\end{abstract}

\section{Introduction}
Microquasars are galactic X-ray binary systems exhibiting relativistic jets \citep{Mirabel94} and are considered in some models to be a possible source of high energy ($>100$~GeV) neutrinos \citep{2001PhRvL..87q1101L,2003A&A...410L...1R}. The composition of the jets and in particular their baryonic content is still an open issue and is a key point for the expectations on the flux of neutrinos. 
Evidence for a baryonic content has been found only in the jets of the microquasars \sss{} and 4U~1630$-$47,
witnessed by the observation of both blue and red Doppler shifted ionization lines of heavy elements \citep{2002ApJ...564..941M,2013arXiv1311.5080D}. Further observations have also indicated reheating (in situ acceleration) of the baryonic component at parsec scales in the jets of \sss{} \citep{2002Sci...297.1673M}. 
Some hints in favour of a significant baryonic content in the outflow of the microquasar \cygxi{} have been inferred using knowledge of the energetics of its jets: the observation of a radio-emitting large structure in the interstellar medium of \cygxi{} has allowed the energy output of the jets to be constrained within the interval $10^{36}-10^{37}\ \mathrm{erg\,s^{-1}}$ \citep{2005Natur.436..819G}, which is two orders of magnitude larger than the estimates based on the jet's flat radio spectrum \citep{2000MNRAS.312..853F}. This excess of energy may be ascribed, among other things, to a population of cold baryons carried in the relativistic flow \citep{2006ApJ...636..316H}. The observation of an X-ray emission from the parsec-scale jets of the microquasars \xtejmdl{} and \hhh{} with \emph{Chandra} may imply the presence of $\sim\!\!10\,\mathrm{TeV}$ electrons, which would be most likely accelerated by shocks in the propagating plasma \citep{2002Sci...298..196C,2005ApJ...632..504C}.

If baryons were actually contained in microquasar jets and a dissipation mechanism allowed them to be accelerated to very high energies, e.g. through diffusive shock acceleration, synchrotron emission from the electrons may provide the required opacity to photo-meson production and high energy neutrinos may be produced \citep{2001PhRvL..87q1101L,2002ApJ...575..378D}. In microquasars harbouring an early type, massive ($\gtrsim 10M_{\odot}$) companion star, neutrinos may be generated by the interaction of the relativistic baryons in the jets with the ions from the stellar wind \citep{2003A&A...410L...1R}. The detection of high energy neutrinos from microquasars would thus give important clues about the composition of microquasar jets and about the physics taking place in the extreme environments close to black holes or neutron stars, and identify microquasars as one of the sources of the galactic component of cosmic rays.

This paper presents a search for neutrino emission from microquasars with the ANTARES detector. In order to maximise the signal to noise ratio, the atmospheric neutrino background is reduced by restricting the data to the times in which jet acceleration is supposed to take place at the sources under study. A multiwavelength approach using X-ray and gamma ray data is applied in order to select the outbursting periods. In \S~\ref{sec:antares}, the ANTARES detector and the data set used in this analysis are described. The selected sources and the criteria adopted to define the time cuts for the neutrino search are presented in \S~\ref{sec:timesel}. In \S~\ref{sec:search}, the statistical method adopted to analyse the data is presented and the results are derived. Conclusions are drawn in \S~\ref{sec:conclusions}.

\section{ANTARES}\label{sec:antares}
The ANTARES neutrino telescope is an underwater detector optimised to detect neutrinos with energies above 100~GeV \citep{2011ApJ...743L..14A,2011PhLB..696...16A}.
High energy neutrinos interacting with the matter surrounding the detector produce relativistic charged particles that induce the emission of Cherenkov light. Among neutrino flavours, muon neutrinos are the favoured probe for astrophysics, since muons' range in water reaches up to several kilometers at the energies of interest for ANTARES, thus allowing a more precise reconstruction of their direction and an effective volume higher than the instrumented volume.

The detector consists of an array of 885 photomultiplier tubes (PMTs) located in the Mediterranean Sea at a depth of 2475~m, 40~km offshore from the Southern coast of France. It is composed of 12 detection lines each hosting 75 optical modules \citep[OMs,][]{2002NIMPA.484..369A}, arranged in 25 storeys, with 3 OMs on each storey. The OMs are inclined towards the bottom by 45$^\circ$ to favour the detection of upward-going tracks. The storeys are equally spaced along the lines by 14.5~m. The spacing between the lines is approximately 60~m.

The light signals detected by the OMs, the ``hits'', are digitised by electronics boards placed on each storey and then sent to a PC farm onshore that performs the filtering of the data \citep{2007NIMPA.570..107A}. The data filtering algorithm is based on the occurrence of ``L1 hits'', i.e., coincidences of two hits within a single storey, or single hits with collected charge higher than 3 photoelectrons. The algorithm selects the events containing at least five causally connected L1 hits or a local cluster of neighbouring L1 hits. An appropriate time calibration procedure is applied to the arrival time of the light signals \citep{2011APh....34..539A}, whereas a positioning system takes care of recording the displacement of the lines, due to the sea currents \citep{2012JInst...7T8002A}. A detailed description of the ANTARES detector and its calibration procedures can be found in \citep[][and references therein]{Antares2011}.

The ANTARES detector started taking physics data in January 2007, when it was composed of only five detection lines. It was upgraded to ten detection lines in December 2007 and was completed in May 2008 with its twelve-line configuration. This paper describes an analysis of the ANTARES data collected between January 2007 and the end of 2010.

The reconstruction of muon tracks is performed by means of a multi-step procedure whose final result is provided by a maximum likelihood fit \citep{2012ApJ...760...53A}. The quality parameter of the reconstructed tracks, here referred to as $\Lambda$, is defined on the basis of the maximised likelihood. The algorithm also provides an estimate of the angular uncertainty of the reconstructed direction event-by-event, here referred to as $\beta$. Only tracks with $\beta<1^\circ$ are selected for the analysis. The response of the detector and the performances of the reconstruction algorithm are estimated by means of Monte Carlo simulations. Atmospheric neutrinos, which represent an irreducible background, are simulated according to the flux obtained by \citet{1996PhRvD..53.1314A}. The flux of atmospheric muons reaching the detector from above is simulated with the code \texttt{MUPAGE} \citep{2008CoPhC.179..915C}.

\begin{figure}
\centering%
\includegraphics[width=0.73\textwidth]{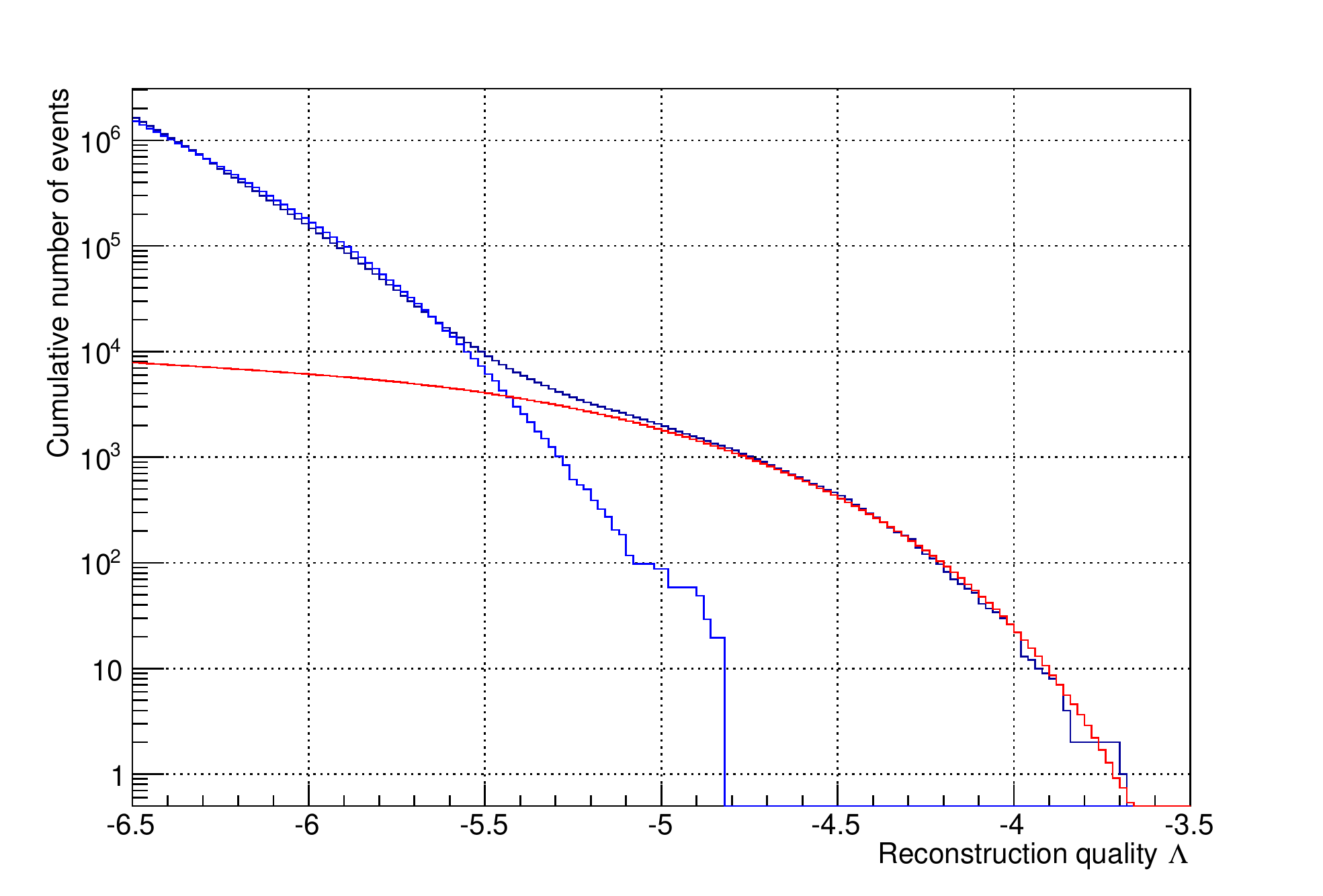}
\caption{Cumulative distribution of the track reconstruction quality parameter $\Lambda$ for the data (black) and for the simulated atmospheric muons (blue) and neutrinos (red), for upgoing tracks with $\beta<1^\circ$.}
\label{fig:lambda_data_mc}
\end{figure}

Due to the intense flux of downgoing atmospheric muons, only tracks reconstructed as upgoing are selected as neutrino candidates. Nevertheless, a fraction of atmospheric muons can be misreconstructed as upgoing and contaminate the neutrino sample. These background tracks are reconstructed with a low quality and can be discarded by applying a suitable cut on $\Lambda$. Two cuts on $\Lambda$ are used in this analysis: $\Lambda>-5.2$ and $\Lambda>-5.4$. These cut values result from the optimisation procedure described in \S~\ref{sec:search}, and lead to a contamination from atmospheric muons of 13\% and 42\%, respectively. \figurename~\ref{fig:lambda_data_mc} shows a comparison between data and Monte Carlo of the distribution of the reconstruction quality $\Lambda$ for upgoing tracks with angular uncertainty $\beta<1^\circ$.

\begin{figure}
\centering%
\includegraphics[width=0.73\textwidth]{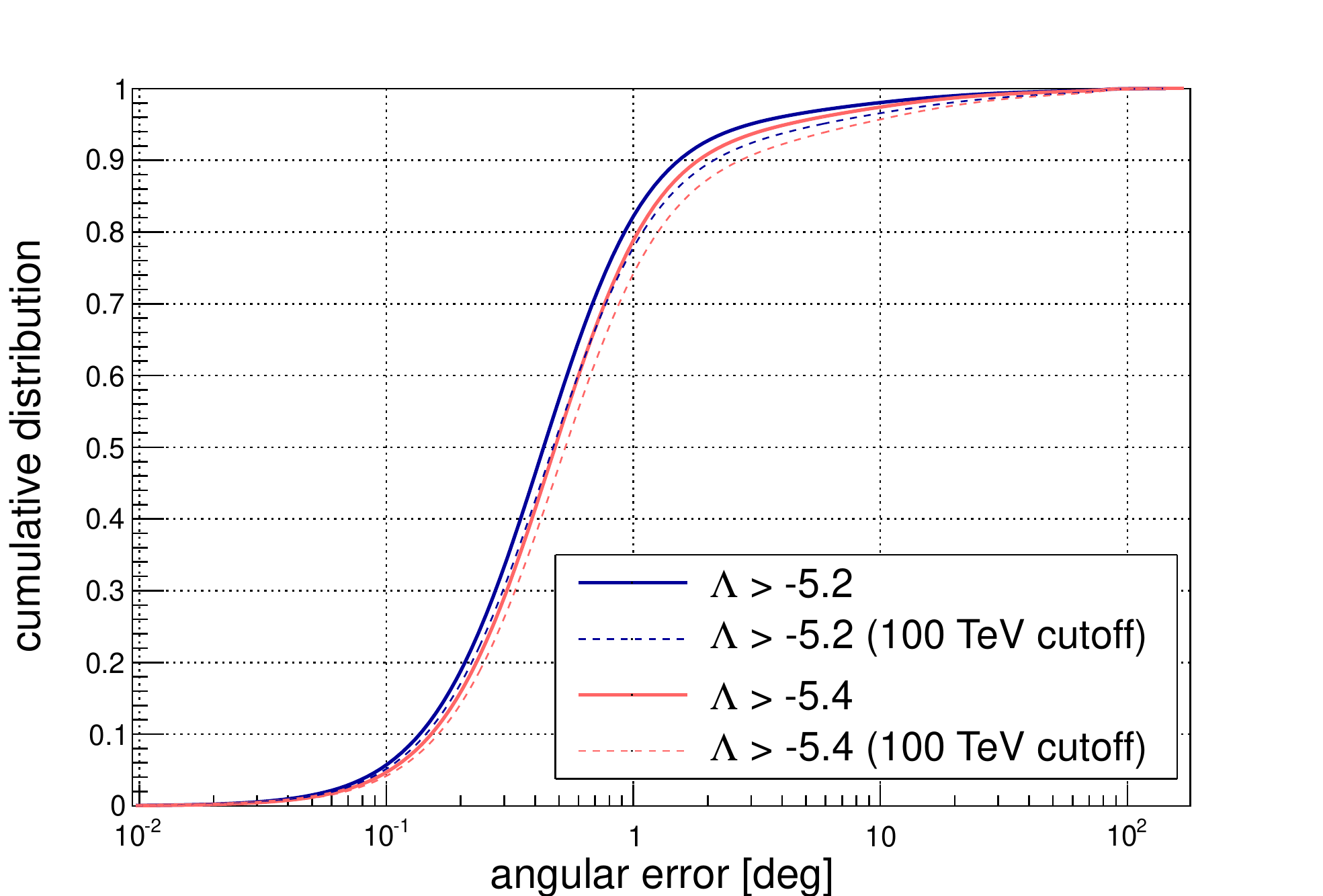}
\caption{Cumulative distribution of the angular error of the reconstruction algorithm for two different quality cuts and a neutrino source with an energy spectrum $\propto E_\nu^{-2}$ (solid) and $\propto E_\nu^{-2}\exp(-\sqrt{E_\nu/100\,\mathrm{TeV}})$ (dashed).}
\label{fig:angres}
\end{figure}

The angular resolution of ANTARES is also estimated using Monte Carlo simulations and depends on the assumed neutrino spectrum. The cumulative distribution of the angular error of the reconstructed neutrinos is shown in \figurename~\ref{fig:angres} for two different quality cuts and a spectrum $\propto E_\nu^{-2}$ and $\propto E_\nu^{-2} \exp(-\sqrt{E_\nu/100\,\mathrm{TeV}})$. The median angular resolution for an $E_\nu^{-2}$ spectrum is $0.46\pm 0.10\,\mathrm{deg}$ \citep{2012ApJ...760...53A}, using a cut $\Lambda>-5.2$.

\section{Selection of flaring time periods}\label{sec:timesel}
Among the sources listed as microquasars in the catalogues of X-ray binaries \citep{Liu:HMXRBCat07,Liu:LMXRBCat07} and visible by ANTARES, those showing an outburst that could be associated with the acceleration of relativistic jets are selected for this analysis. The selected sources, ordered by increasing declination, are: \cirxi{}, \gx{}, \hhh{}, \igrj{}, \cygxi{} and \cygxiii{}. To maximise the detection probability, the search for coincident neutrino events is restricted to microquasar flaring time periods. The selection of flaring periods is carried out by taking into account the multiwavelength behaviour of each of the selected microquasars, using information from X-ray or gamma-ray instruments.

For all the microquasars considered in this paper, except \cygxiii{}, the time selection is based on their X-ray behaviour. When the procedure involves the selection of X-ray outbursts in the daily averaged X-ray light curves of RXTE/ASM\footnote{RXTE/ASM data are taken from http://xte.mit.edu/asmlc/} (1.2--12 keV) or Swift/BAT\footnote{Swift/BAT data are taken from http://heasarc.nasa.gov/docs/swift/results/transients/} (15--50 keV), a dedicated outburst selection algorithm is adopted.

To infer the onset of an X-ray outburst, a Gaussian fit is performed to the distribution of the X-ray rates to calculate the baseline rate, $\langle r \rangle$, and its standard deviation, $\sigma_{\langle r \rangle}$. The fit is iterated twice within the interval $\langle r \rangle_{-2\sigma_{\langle r \rangle}}^{+\sigma_{\langle r \rangle}}$, to reduce the contribution from the bursting region. The light curve is then scanned, looking for a rate measurement $\tilde{r}$ satisfying the condition $\tilde{r} - 3\sigma_{\tilde{r}}> \langle r \rangle+3\sigma_{\langle r \rangle}$, where $\sigma_{\tilde{r}}$ is the error on the single rate measurement. Once such a measurement is found, all the following and preceding measurements for which $\tilde{r} - \sigma_{\tilde{r}}>\langle r \rangle+3\sigma_{\langle r \rangle}$ are selected. At least two consecutive measurements are required for the selection to take place. In the case of \cygxiii{} the time selection is based on the Fermi/LAT gamma-ray light curve, hence a slightly different procedure is used to select the outbursts, as described in \S~\ref{sec:cygx3}.

As the increase of the X-ray flux alone is not generally sufficient to ensure the onset of a relativistic jet, additional time selection criteria, customised for the features of each microquasar under study, are applied and are described in the following.

\subsection{Black hole binaries}
Four of the microquasars considered in this analysis are black hole candidates or confirmed black hole binaries. The time evolution of X-ray outbursts in this type of source is known to follow a specific pattern in the intensity and spectrum of the X-ray flux \citep[see][for a review]{2010LNP...794...53B}. The beginning of an outburst is characterised by a power-law energy spectrum with photon index $\sim 1.7$, known as the \emph{hard state}, during which a steady jet is observed with Lorenz factor $\sim 2$. This is followed by an X-ray state in which the hard power law component is almost suppressed in favour of a soft thermal component with temperature $\sim 1$ keV, referred to as the \emph{soft state}, during which the radio jet is suppressed and the disk emission is dominant. The transition between these two canonical states, which are rather stable and can last several weeks, takes place through two intermediate states, the hard and soft intermediate states (HIMS and SIMS, respectively), both characterised by spectral features between the hard and soft states. These transitions occur on time scales of hours/days and are often associated with discrete-ejection observed in radio wavelengths whose Lorenz factor is thought to be higher than that observed during the hard state \citep{2004MNRAS.355.1105F,2009MNRAS.396.1370F}.

The time selection for the following sources is based entirely on their X-ray behaviour, by taking into account the disk-jet coupling just described. Only periods of hard X-ray states and state transitions are selected, since they correspond to phases where relativistic jets are present, and therefore neutrino emission is expected. Also, both states are considered separately for the subsequent neutrino search.

\subsubsection{\gx{}}
\gx{} is a galactic black hole binary system \citep{2003ApJ...583L..95H,2004MNRAS.351..791Z} that has undergone two major outbursts between 2007 and 2010, both  featuring a hard to soft transition, and some fainter ones (see \figurename~\ref{fig:lightcurves}).

The hard states have been selected as outbursts in the Swift/BAT daily averaged light curve, using the procedure described at the beginning of this section. The times of the transitions as well as of the onset of the soft state are estimated using X-ray spectral observations, and are used to define the end of the hard state.

At the beginning of 2007 the source was already in outburst. From a detailed study of the spectral time evolution of \gx{} during this outburst, the transition from the HIMS to the SIMS is observed around MJD~54145.5, and then again around MJD~54160 and MJD~54164 \citep{2009MNRAS.400.1603M}. A similar outburst was again observed in the first half of 2010. The decay phase of this outburst and the subsequent transition to softer states has been followed by RXTE/PCA pointed observations by \citet{ATel_2577}, who locate the transition from the HIMS to the SIMS at around MJD~55304, after which the source was observed undergoing a transition to the soft state and then to the SIMS until MJD 55316. On MJD~55320 the X-ray spectrum was compatible with the source being in the HIMS again \citep{ATel_2593}, but the subsequent transition was not observed, though from the light curves and the similarity with the 2007 outburst, the time at which it occurred can be estimated to be around MJD 55324. A time window of 5 days centred at the estimated time of the state transitions is selected for the neutrino search. Its start time coincides with the end of the hard state period preceding it.

\subsubsection{\hhh{}}
For the black hole candidate \hhh\ \citep{2009ApJ...698.1398M} the same selection procedure used for \gx{} is adopted. Between 2007 and 2010, \hhh\ has undergone five outbursts (\figurename~\ref{fig:lightcurves}).

The outburst that occurred at the end of 2008 has been classified as a failed outburst \citep{2009MNRAS.398.1194C}, since the source remained between the hard state and the HIMS without reaching the soft state or the SIMS. The subsequent outburst that occurred in 2009 featured, after the onset of the hard state, a HIMS$\rightarrow$SIMS transition at MJD 54990 \citep{2010A&A...522A..99C}, which is included in a time window of  $\pm$2.5 days for the transitional phase search. The evolution of the outburst detected in 2010 is described in \citep[and references therein]{ATel_2774}. A HIMS$\rightarrow$SIMS transition is observed around MJD~55424.5, after which the source stays in the soft state until quiescence. Two more outbursts were observed while the source was close to the Sun and thus received a poor coverage by X-ray telescopes. The first was detected around MJD 54453 and lasted until MJD 54504 \citep{ATel_1348,2010MNRAS.401.1255J}. The second was detected around MJD 55191 until MJD 55237 \citep[][and references therein]{ATel_2364}. These are also included in the analysis as hard states, since it is not known whether state transitions have taken place.

\subsubsection{\igrj{}}
For the black hole candidate \igrj{} \citep{2006A&A...446..471P} a short outburst was observed in the Swift/BAT  daily light curve between MJD~54286 and MJD~54301 (\figurename~\ref{fig:lightcurves}), which has been selected for the analysis. Swift/XRT observations during that period confirmed a hardening of the X-ray spectrum \citep{ATel_1140}.

\subsubsection{\cygxi{}}
The time selection for the black hole binary \cygxi{} \citep{1965Sci...147..394B,1972Natur.235...37W,2001MNRAS.327.1273S} is performed differently to the black hole binaries or black hole candidates discussed previously. By comparing simultaneous RXTE/PCA and RXTE/ASM data on a long term monitoring of \cygxi{}, \citet{2013A&A...554A..88G} have shown how to perform an almost exact mapping of the X-ray spectral state on the basis of the sole RXTE/ASM flux and hardness.  
This mapping is used in order to define the onset of a hard or intermediate state for this source. The results of this selection are shown in \figurename~\ref{fig:lightcurves}.

\subsection{\cirxi}
\cirxi{} is the only confirmed neutron star microquasar considered in this analysis. It has an orbital period of 16.6 days \citep{1976ApJ...208L..71K} and undergoes regular radio flares with the same period, interpreted as enhanced accretion near periastron passage in a highly eccentric orbit \citep{1980A&A....87..292M}. A high angular resolution monitoring campaign in the radio, conducted with e-VLBI in 2009, has confirmed this behaviour, observing an enhanced radio emission between orbital phase 0.09 and 0.21 \citep{2011MNRAS.414.3551M}, although these measurements were taken during a period of very low activity in both radio and X-rays. Simultaneous radio and X-ray observations have shown that an increased accretion rate, in the form of an X-ray outburst, is followed by the acceleration of relativistic jets, observed as a brightening of the radio core of the source, followed by the brightening of the arcsecond scale radio structure surrounding the core \citep{2004Natur.427..222F}.

This source has undergone several X-ray outburst events between 2007 and 2010, better visible in the soft X-rays (\figurename~\ref{fig:lightcurves}), that are included in the analysis using the procedure described at the beginning of this section applied to the daily averaged light curves of RXTE/ASM. If sufficiently close in time, i.e. within 4 days, the selected periods are extended, forward or backward, in order to include the expected radio flare at superior conjunction.

\subsection{\cygxiii{}}\label{sec:cygx3}
\cygxiii{} is a high-mass binary \citep{1992Natur.355..703V,1972Natur.239..123P} in which the nature of the compact object has not yet been identified. It has been observed emitting high energy gamma-rays by both AGILE \citep{2009Natur.462..620T} and Fermi/LAT \citep{2009Sci...326.1512F}, in association with its ultra soft X-ray states which in turn are related to giant radio outbursts. Thus the time selection for this source is based on its behaviour in gamma-rays.

\begin{figure}
\centering%
\includegraphics[width=0.7\textwidth]{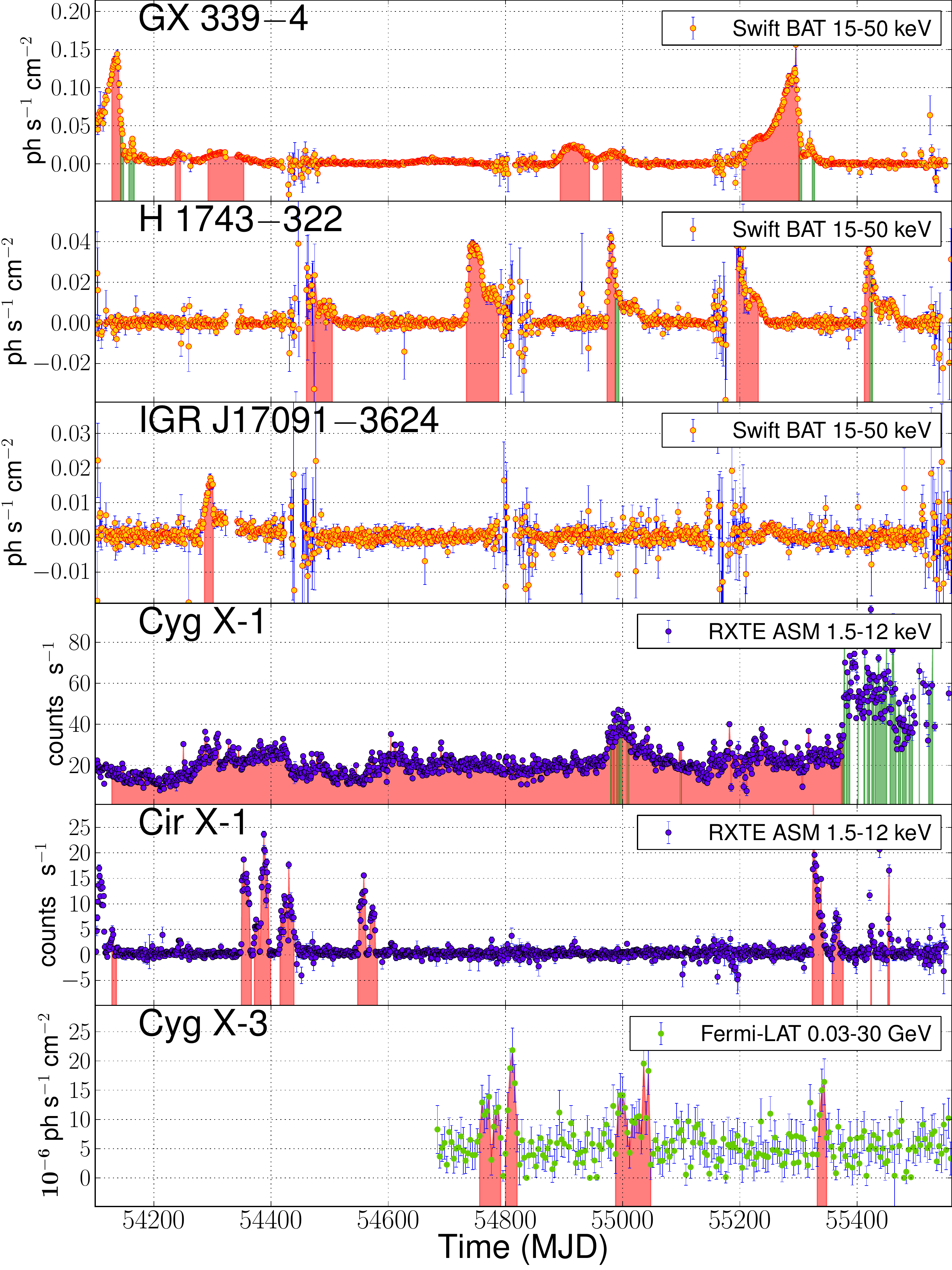}
\caption{X-ray and gamma-ray light curves (Swift/BAT in red, RXTE/ASM in blue and Fermi/LAT in green) used for the selection of the flaring times. The shaded areas represent the times selected for the analysis. The red and green areas in the light curves of \gx{}, \hhh{} and \cygxi{} correspond to hard states and state transitions, respectively.}
\label{fig:lightcurves}
\end{figure}

The gamma-ray data between 30~MeV and 30~GeV from Fermi/LAT\footnote{Fermi/LAT data have been retrieved from the web page http://fermi.gsfc.nasa.gov/cgi-bin/ssc/LAT/LATDataQuery.cgi.} are analysed using the procedure described by \citet{2009Sci...326.1512F}\footnote{This analysis has been performed using \texttt{HEASOFT v6.11}, \texttt{ScienceTools v9r23p1} and the response function \texttt{P6\_V1\_DIFFUSE}.}. The light curve is calculated in time bins of four days and is shown in \figurename~\ref{fig:lightcurves}. The selection of gamma-ray flares is performed in the same fashion as the one described for the X-rays at the beginning of this section, with some modifications due to the different nature of the data. Namely, a Gaussian fit is performed on the distribution of the gamma-ray rate $r$, to calculate its mean value $\langle r \rangle$ and its standard deviation $\sigma_{\langle r \rangle}$. All the times corresponding to a flux measurement $\tilde{r}$ for which $\tilde{r}-\sigma_{\tilde{r}}> \langle r \rangle$, where $\sigma_{\tilde{r}}$ is the error on the flux measurement, are then included in the analysis within a time window of 4 days, which is the binning used to produce the light curve. The obtained time windows are extended by 5 days before and after, to take into account the possible time lags between the gamma-ray emission and the development of the jets \citep{2011ApJ...733L..20W}. The results of this selection are shown in \figurename~\ref{fig:lightcurves}.

\begin{table}
\centering%
\caption{Candidate microquasars and selected periods.}
\label{tab:timesel}
\scriptsize%
\begin{tabular}{lc}
\toprule
    Source Name    &  Selected periods (MJD)  \\
\midrule
\cirxi{} &  54128-54136, 54349-54367, 54371-54400, 54415-54442, 54547-54582, 55323-55343, 55357-55375,\\
              &  55421-55426, 55452-55458 \\
\midrule
\gx{}     & 54128-54143, 54236-54246, 54292-54354, 54893-54944, 54966-54998, 55203-55301\\
(hard state) & \\
\midrule
\gx{}     &  54143-54148, 54157.5-54166.5, 55301-55306, 55323-55328\\
(transition) &    \\
\midrule
\hhh{}    &  54453-54504, 54733-54789, 54973-54987.5, 55191-55237, 55412-55422\\
(hard state) &    \\
\midrule
\hhh{}    &  54987.5-54992.5, 55422-55427  \\
(transition) &   \\
\midrule
\igrj{}      &  54286-54301\\
(hard state) &              \\
\midrule
\cygxi{}     &  54128-54979, 54980-54984, 54985-54990, 54991-54992, 54993-54995, 54997-54998,  54999-55007,\\
(hard state) &   55008-55009,55010-55099, 55100-55374, 55375-55377\\
             & \\
             &  \\
\midrule
\cygxi{}    & 54979-54980, 54984-54985, 54990-54991, 54992-54993, 54995-54997, 54998-54999, 55007-55008,\\
(transition) &  55009-55010,55099-55100, 55374-55375, 55377-55379, 55381-55388, 55401-55402, 55411-55414,\\
             &  55417-55418, 55419-55422, 55425-55429, 55430-55451, 55456-55467, 55470-55475, 55477-55483,\\
             &    55484-55485, 55488-55492, 55494-55495, 55506-55507, 55522-55523, 55528-55530 \\
\midrule
\cygxiii{}   &   54753-54795, 54797-54823, 54985-55051, 55329-55351 \\
\bottomrule
\end{tabular}
\end{table}

\section{Search for coincident neutrino events}\label{sec:search}
\subsection{Statistical method}
\begin{figure}
\centering%
\includegraphics[width=0.60\textwidth]{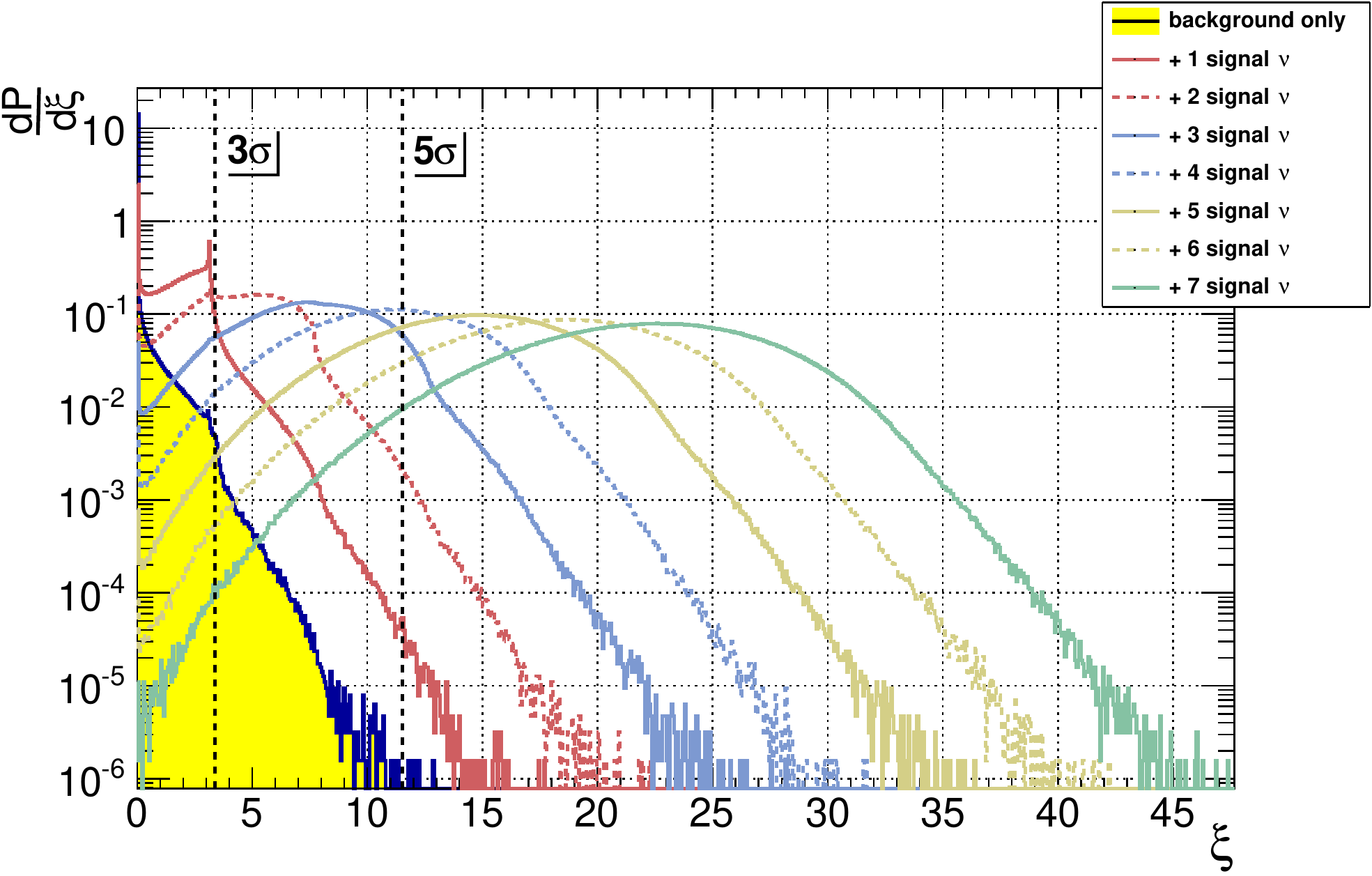}
\caption{Probability distribution of the test statistic variable issued from pseudo experiments for background only (solid histogram) and by adding from 1 up to 7 signal neutrinos around the source. The vertical dashed lines indicate the threshold values of $\xi$ that lead to a $3\sigma$ and $5\sigma$ pre-trial rejection of the background only hypothesis. This plot corresponds to the case of \gx{} during hard states.}
\label{fig:tsdistro}
\end{figure}

\begin{figure}
\centering%
\includegraphics[width=0.60\textwidth]{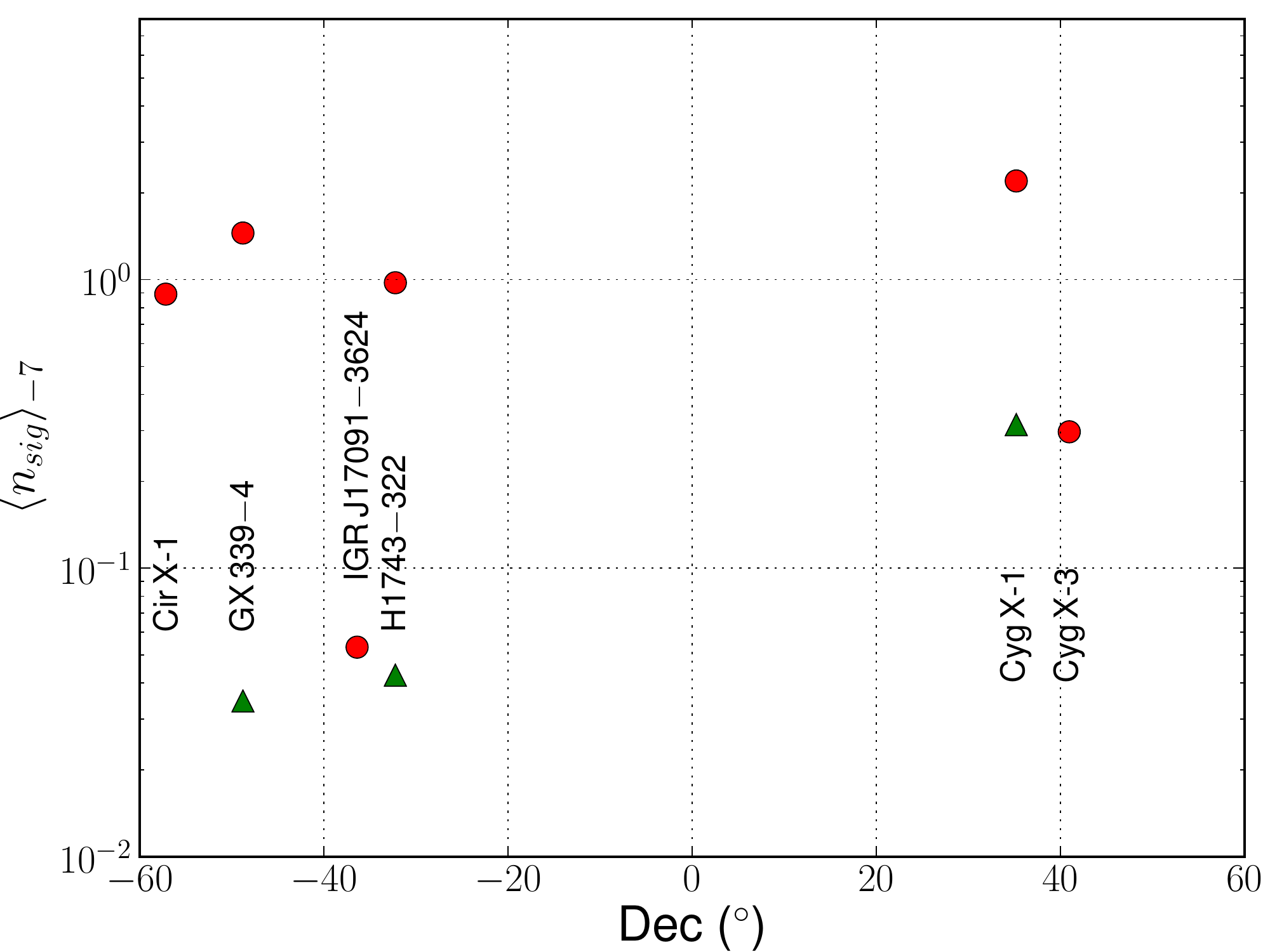}
\caption{Number of neutrino events expected by ANTARES from a source with an $E_\nu^{-2}$ spectrum and a flux normalisation of $10^{-7} \mathrm{GeV\,cm^{-2}\,s^{-1}}$ when applying the time and quality cuts listed in \tablename~\ref{tab:timesel} and \tablename~\ref{tab:results}, as a function of the source declination. The green triangles stand for black hole binary transitional states.}
\label{fig:acceptance}
\end{figure}

The ANTARES data collected between 2007 and 2010, corresponding to 813 days of livetime, are analysed to search for neutrino events around the selected sources, in coincidence with the time periods defined in the previous section and listed in \tablename~\ref{tab:timesel}. The statistical method adopted to infer the presence of a signal on top of the atmospheric neutrino background, or alternatively set upper limits on the neutrino flux is an unbinned method based on a likelihood ratio test statistic. The likelihood is defined as:
\begin{equation}
\label{eq:unbinlik}
\log(\mathcal{L})=\sum_{i=1}^{n_{\rm tot}}\log[n_{\rm sig}\mathcal{S}(\alpha_i)+n_{\rm tot}\mathcal{B}(\theta_i)] - n_{\rm tot}\ ,
\end{equation}
where $n_{\rm tot}$ is the total number of neutrino events detected during the flaring periods and while the source was visible by ANTARES (i.e., below the horizon). In \eq~\ref{eq:unbinlik}, $\mathcal{S}$ is the point spread function (PSF), $\alpha_i$ is the angular distance of the event $i$ from the position of the source, $\mathcal{B}$ is the distribution of background events as a function of the zenith angle $\theta_i$, and both $\mathcal{S}$ and $\mathcal{B}$ are normalised to 1. 
A spectrum of the form $\dd N/\!\!\dd E_\nu=\phi E_\nu^{-2}\,\mathrm{GeV\, cm^{-2}\, s^{-1}}$ is used to optimise the search and to calculate the upper limits of the neutrino fluences, whereas a customised spectral shape is used to compare the results with model predictions (\S~\ref{sec:results}). The normalisation constant $\phi$ is the quantity to be measured or upon which upper limits are set. 
The result of each search is based on the value assumed by the test statistic variable $\xi$, which is defined as the logarithm of the ratio between the likelihoods calculated under the hypotheses of background plus signal and background-only:
\begin{equation}
\xi=\max\{\log\mathcal{L}(n_{\rm sig})\} - \log\mathcal{L}(n_{\rm sig}=0)\, .
\end{equation}
Monte Carlo pseudo experiments are generated to compute the distributions of $\xi$ under the background only and background plus signal hypotheses. 
Each pseudo experiment simulates the number of neutrinos selected in the data. The simulated neutrino directions for the background events are randomly generated according to the zenith and azimuth distribution of the neutrinos selected in the whole ANTARES 2007-2010 data set. The conversion to celestial coordinates is done using the true time of the detected event. To simulate the presence of a signal, pseudo experiments are also generated by adding from one up to thirty neutrinos distributed around the source according to the ANTARES PSF. An example of the resulting distributions of the test statistic for background-only and background plus a fixed number of injected signals, $P(\xi|n_{\rm sig})$, is shown in \figurename~\ref{fig:tsdistro} for the case of \gx{} outbursts during hard states. The distribution of the test statistic for a mean number of signals $P(\xi|\langle n_{\rm sig} \rangle)$ is calculated from a Poissonian convolution of the $P(\xi|n_{\rm sig})$:
\begin{equation}
P(\xi|\langle n_{\rm sig} \rangle) = \sum_{n_{\rm sig}} P(\xi|n_{\rm sig})\frac{\langle n_{\rm sig}\rangle^{n_{\rm sig}}e^{-\langle n_{\rm sig}\rangle}}{n_{\rm sig}!}\,,
\end{equation}
and is used to construct the 90\% confidence belts with the unified approach of \citet{Feldman_Cousins_1997}. The conversion between $\langle n_{\rm sig} \rangle$ and the normalisation of the neutrino flux $\phi$ is performed by means of Monte Carlo simulations.  The average number of expected neutrino events $\langle n_{\rm sig} \rangle_{-7}$ from the selected sources assuming an $E_\nu^{-2}$ spectrum and a flux normalisation $\phi_{-7}=10^{-7}\,\mathrm{GeV\,cm^{-2}\,s^{-1}}$ is shown in \figurename~\ref{fig:acceptance} as a function of the source declination. This quantity is used to convert the upper limits $\langle n_{\rm sig} \rangle^{90\%\,\rm CL}$ to an upper limit on the flux normalisation $\phi^{90\%\,\rm CL}$ by a proportional scaling:
\begin{equation}
\phi^{90\%\,\rm CL}=\phi_{-7}\frac{\langle n_{\rm sig} \rangle^{90\%\,\rm CL}}{\langle n_{\rm sig} \rangle_{-7}}\,.
\end{equation}
The quality cuts on $\Lambda$ to be used in each neutrino search are optimised in order to minimise the flux needed for a $5\sigma$ discovery in $50\%$ of pseudo experiments and are listed in \tablename~\ref{tab:results}. The optimisation is carried out only on the basis of the results of the pseudo experiments and while keeping the true neutrino directions in the data unknown.

\subsection{Results}\label{sec:results}
The above statistical method has been applied to calculate the $\xi$ value of each neutrino search. The results are summarised in \tablename~\ref{tab:results}. As none of the searches has produced a statistically significant neutrino excess above the expected background, the 90\% confidence level upper limits on the flux normalisation of an $E_\nu^{-2}$ spectrum, $\phi^{\mathrm{90\%\rm CL}}$, are calculated. Systematic uncertainties of 15\% on the angular resolution and 15\% on the detector acceptance have been included in the upper limit calculations. These systematic errors have been constrained on the basis of a 30\% uncertainty on the atmospheric neutrino flux as shown by \cite{2012ApJ...760...53A}. Also, a systematic uncertainty on the absolute orientation of the detector of $\sim 0.1\,\mathrm{deg}$ has been taken into account \citep{2012JInst...7T8002A}. The $\phi^{\mathrm{90\%\rm CL}}$ are used to obtain the upper limits of the neutrino fluences, i.e. the energy per unit area, as:
\begin{equation}
\mathcal{F}_\nu^{90\%\,CL} = \phi^{90\%\,\rm CL}\Delta T_{\rm search} \int_{10^2\:\mathrm{GeV}}^{10^8\:\mathrm{GeV}}E_\nu\cdot E_\nu^{-2}\dd E_\nu\,,
\end{equation}
where $\Delta T_{\rm search}$ is the corresponding livetime of the search. The values obtained are reported in \tablename~\ref{tab:results}.

In order to compare the results with the expectations from the model by \cite{2001PhRvL..87q1101L} and reported by \cite{2002ApJ...575..378D}, the upper limits on the flux normalisation are also calculated considering a neutrino spectrum $\propto E_\nu^{-2}\exp(-\sqrt{E_\nu /100\,\mathrm{TeV}})$, i.e., with an exponential cutoff at 100~TeV to take into account the limitation in the acceleration process included in the model. \citet{2002ApJ...575..378D} express their results in terms of the energy flux of neutrinos $f_\nu = \mathcal{F}_\nu/\Delta T_{\rm search}$ and with respect to their calculation a factor 0.5 is applied here to account for muon neutrino disappearance due to neutrino oscillations, which was not included in their paper. The model prediction for \hhh{} is not given by \cite{2002ApJ...575..378D} and is calculated using the near-infrared observation by \cite{2003IAUC.8112....B} during the 2003 outburst. \cite{2003IAUC.8112....B} detected a magnitude 13.6 in the 2MASS K$_s$-band, corresponding to a flux density of 2.4~mJy at a frequency of $1.4\times10^{14}\ \rm Hz$, which allows the calculation of the model expectation of the energy flux $f_\nu$ using \eq~8 in \cite{2002ApJ...575..378D}. No prediction for \igrj{} is given by \citet{2002ApJ...575..378D}, nor measurements were found to estimate it.

\begin{figure}
\centering%
\includegraphics[width=0.70\textwidth]{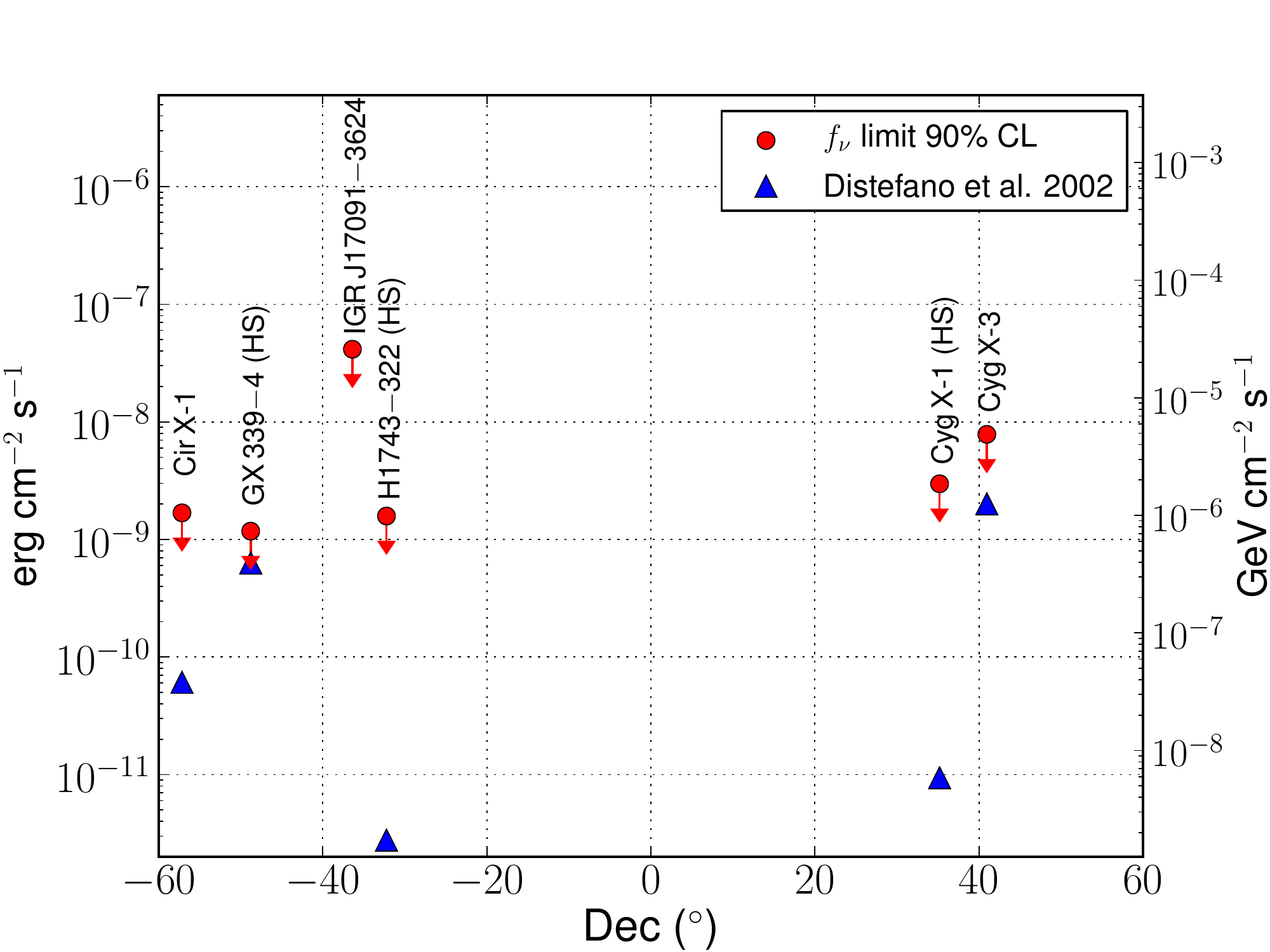}
\caption{Feldman-Cousins 90\% confidence level upper limits on the energy flux in neutrinos $f_\nu$ obtained in this analysis considering a flux  $\propto E^{-2}\exp(-\sqrt{E_\nu/100\ \rm TeV})$ (circles), compared with the expectations by \cite{2002ApJ...575..378D} (triangles).}
\label{fig:fnucomparison}
\end{figure}

The 90\% confidence level upper limit on the energy flux of neutrinos obtained from this analysis and the corrected values of the expectations from \cite{2002ApJ...575..378D} are reported in \tablename~\ref{tab:results} and shown in \figurename~\ref{fig:fnucomparison}. All our results are above the expectations and thus no constraints can be put on the model parameters for any of the sources. The limit for \gx{}, which is one the most promising sources according to \cite{2002ApJ...575..378D} calculations, is a factor $\sim 2$ above the expectations. Depending on the \gx{} outbursting duty cycle in the future, additional ANTARES data may allow to reach a sufficient sensitivity to eventually constrain some of the model parameters.

\begin{table}[t!]
\centering%
\scriptsize%
\begin{tabular}{lccccccccc}
\toprule
Source         & $\Lambda>$  & $\xi$ & livetime & $n_{\rm tot}$ & closest $\nu$ & $\mathcal{F}_\nu^{90\% \mathrm{CL}}$  &  \,  &   $f_\nu^{90\% \mathrm{CL}}$   &   $f_\nu^{th}$\\
               &             &       &  (days)     &           &              &   $(\mathrm{GeV\,cm^{-2}})$          &  \,  &   $(\mathrm{erg\,cm^{-2}\,s^{-1}})$  &    $(\mathrm{erg\,cm^{-2}\,s^{-1}})$ \\
\midrule
\cirxi         & $-5.2$  &   0     &  100.5  & 257  &  5.7$^\circ$  &  16.8   &  \,   &  $1.68\times 10^{-9}$    &   $6.10\times 10^{-11}$   \\
\gx{}  (HS)     & $-5.2$  &   0     &  147.0  & 485  &  2.8$^\circ$  &  10.9   &  \,   &  $1.18\times 10^{-9}$    &   $6.30\times 10^{-10}$   \\
\gx{}  (TS)     & $-5.4$  &   0     &   4.9   &  14  &  11$^\circ$  &   19.4  &  \,   &  $6.27\times 10^{-8}$    &   $6.30\times 10^{-10}$   \\
\hhh{} (HS)    & $-5.2$  &   0     &   83.6  & 444  &  4.6$^\circ$  &   9.2   &  \,   &  $1.58\times 10^{-9}$    &   $2.78\times 10^{-12}$   \\
\hhh{} (TS)      & $-5.4$  &   0     &   3.3   & 22   & 15.9$^\circ$  &   10.2  &  \,   &  $4.33\times 10^{-8}$    &   $2.78\times 10^{-12}$   \\
\igrj          & $-5.4$  &   0     &   8.5   &  40  &  12$^\circ$  &  21.0   &  \,   &  $4.15\times 10^{-8}$    &    $-$  \\
\cygxi{} (HS)  & $-5.2$  &   0     &  182.8  &  671 &  1.4$^\circ$  &   9.4   &  \,   &  $2.98\times 10^{-9}$    &   $9.40\times 10^{-12}$   \\
\cygxi{} (TS)  & $-5.4$  &   0     &   18.5  &  117 &  6.4$^\circ$  &  6.0    &  \,   &  $6.75\times 10^{-9}$    &   $9.40\times 10^{-12}$   \\
\cygxiii       & $-5.4$  &   0     &   16.6  & 144  &  6.9$^\circ$  &  5.7    & \,    &  $7.83\times 10^{-9}$    &   $2.01\times 10^{-9}$    \\
\bottomrule
\end{tabular}
\caption{Summary of the results of the neutrino searches for the outbursting microquasars under study. The columns report the values of the adopted cut on the track reconstruction quality $\Lambda$, the test statistic $\xi$, the livetime of the search, the number of neutrinos selected in the whole sky during the selected periods and while the source was below the horizon, the distance of the closest of these neutrinos to the source, the 90\% C.L. upper limit on the neutrino fluence for an $E_\nu^{-2}$ spectrum, the 90\% C.L. of the energy flux of neutrinos for an $E_\nu^{-2}\exp(-\sqrt{E_\nu/100\,\mathrm{TeV}})$ spectrum and the corresponding expectation $f_\nu^{th}$ from \cite{2002ApJ...575..378D}, respectively.}
\label{tab:results}
\end{table}

\section{Conclusions}\label{sec:conclusions}
This paper presents a search for neutrino emission from microquasars during outbursts with the data collected by the ANTARES telescope between 2007 and 2010. The search has been performed under the hypothesis that relativistic jets from microquasars contain baryons that interact during their acceleration or propagation, and time cuts have been chosen to isolate jet acceleration events. The searches did not result in a statistically significant excess above the expected background, thus the $90\%\,\mathrm{C.L.}$ upper limits on the neutrino fluences have been calculated. The results have been compared to the expectations of the neutrino energy flux from a theoretical model and  the obtained upper limits are above the model predictions. The measured flux upper limits are within factors 2 and 4 of the predicted fluxes for the sources \gx{} and \cygxiii{}, respectively. This offers the prospect that they be detectable in the near future either with additional ANTARES data or, in the longer term, by the forthcoming KM3NeT neutrino telescope.

\noindent \textbf{Acknowledgements}


The authors acknowledge the financial support of the funding agencies:
Centre National de la Recherche Scientifique (CNRS), Commissariat \`a
l'\'ene\-gie atomique et aux \'energies alternatives (CEA), Agence
National de la Recherche (ANR), Commission Europ\'eenne (FEDER fund
and Marie Curie Program), R\'egion Alsace (contrat CPER), R\'egion
Provence-Alpes-C\^ote d'Azur, D\'e\-par\-tement du Var and Ville de La
Seyne-sur-Mer, France; Bundesministerium f\"ur Bildung und Forschung
(BMBF), Germany; Istituto Nazionale di Fisica Nucleare (INFN), Italy;
Stichting voor Fundamenteel Onderzoek der Materie (FOM), Nederlandse
organisatie voor Wetenschappelijk Onderzoek (NWO), the Netherlands;
Council of the President of the Russian Federation for young
scientists and leading scientific schools supporting grants, Russia;
National Authority for Scientific Research (ANCS), Romania; Ministerio
de Ciencia e Innovaci\'on (MICINN), Prometeo of Generalitat Valenciana
and MultiDark, Spain; Agence de l'Oriental and CNRST, Morocco. We also
acknowledge the technical support of Ifremer, AIM and Foselev Marine
for the sea operation and the CC-IN2P3 for the computing facilities.

\bibliographystyle{aa}

\input{uqso_paper-arxiv.bbl}
\end{document}

%% file: antares_authors_aastex.tex
\author{
S.~Adri\'an-Mart\'inez\altaffilmark{1},
A.~Albert\altaffilmark{2},
M.~Andr\'e\altaffilmark{3},
M.~Anghinolfi\altaffilmark{4},
G.~Anton\altaffilmark{5},
M.~Ardid\altaffilmark{1},
T.~Astraatmadja\altaffilmark{6},
J.-J.~Aubert\altaffilmark{7},
B.~Baret\altaffilmark{8},
J.~Barrios-Mart\'{\i}\altaffilmark{9}
% ...
}
\altaffiltext{1}{\scriptsize{Institut d'Investigaci\'o per a la Gesti\'o Integrada de les Zones Costaneres (IGIC) - Universitat Polit\`ecnica de Val\`encia. C/  Paranimf 1 , 46730 Gandia, Spain}}
\altaffiltext{2}{\scriptsize{GRPHE - Institut universitaire de technologie de Colmar, 34 rue du Grillenbreit BP 50568 - 68008 Colmar, France}}
\altaffiltext{3}{\scriptsize{Technical University of Catalonia, Laboratory of Applied Bioacoustics, Rambla Exposici\'o,08800 Vilanova i la Geltr\'u,Barcelona, Spain}}
\altaffiltext{4}{\scriptsize{INFN - Sezione di Genova, Via Dodecaneso 33, 16146 Genova, Italy}}
\altaffiltext{5}{\scriptsize{Friedrich-Alexander-Universit\"at Erlangen-N\"urnberg, Erlangen Centre for Astroparticle Physics, Erwin-Rommel-Str. 1, 91058 Erlangen, Germany}}
\altaffiltext{6}{\scriptsize{Nikhef, Science Park,  Amsterdam, The Netherlands}}
\altaffiltext{7}{\scriptsize{CPPM, Aix-Marseille Universit\'e, CNRS/IN2P3, Marseille, France}}
\altaffiltext{8}{\scriptsize{APC, Universit\'e Paris Diderot, CNRS/IN2P3, CEA/IRFU, Observatoire de Paris, Sorbonne Paris Cit\'e, 75205 Paris, France}}
\altaffiltext{9}{\scriptsize{IFIC - Instituto de F\'isica Corpuscular, Edificios Investigaci\'on de Paterna, CSIC - Universitat de Val\`encia, Apdo. de Correos 22085, 46071 Valencia, Spain}}
% ...
